\documentclass[conference]{IEEEtran}
\usepackage{fancyhdr}
\IEEEoverridecommandlockouts
\usepackage{cite}
\usepackage{amsmath,amssymb,amsfonts}
\usepackage{algorithmic}
\usepackage{algorithm}
\usepackage{graphicx}
\usepackage{textcomp}
\usepackage{xcolor}
\usepackage{booktabs}
\usepackage{multirow}
\usepackage{url}
\usepackage[hidelinks]{hyperref}
\usepackage{listings}

\usepackage{tikz}
\usetikzlibrary{shapes.geometric, arrows.meta, positioning, fit, backgrounds, calc}

\newif\ifdraft

 \draftfalse

\def\BibTeX{{\rm B\kern-.05em{\sc i\kern-.025em b}\kern-.08em
    T\kern-.1667em\lower.7ex\hbox{E}\kern-.125emX}}

\begin{document}

\title{Agentic MLIR-to-QIR Translation with Dual-Backend Testing: A Hybrid Deterministic and LLM Approach
\thanks{This manuscript has been authored by UT-Battelle, LLC, under Contract No.\ DE-AC05-00OR22725 with the U.S.\ Department of Energy. The United States Government retains and the publisher, by accepting the article for publication, acknowledges that the United States Government retains a non-exclusive, paid-up, irrevocable, world-wide license to publish or reproduce the published form of this manuscript, or allow others to do so, for the United States Government purposes. The Department of Energy will provide public access to these results of federally sponsored research in accordance with the DOE Public Access Plan.}}
\author{
\IEEEauthorblockN{Sharmin Afrose, Vicente Leyton-Ortega, Narasinga Rao Miniskar, Elaine Wong, Travis S. Humble}
\IEEEauthorblockA{\textit{Oak Ridge National Laboratory}, Oak Ridge, TN, USA \\
\{afroses, leytonorteva, miniskarnr, wongey, humblets\}@ornl.gov}
}

\maketitle

\thispagestyle{fancy} \lhead{} \rhead{} \chead{} \lfoot{\footnotesize{ SC26 Workshops, November 15-20, 2026, Chicago, Illinois, USA
\newline 979-8-3195-1221-5/26/\$31.00 \copyright 2026 IEEE}} \rfoot{} \cfoot{} \renewcommand{\headrulewidth}{0pt} \renewcommand{\footrulewidth}{0pt}

\begin{abstract}
Quantum frameworks each define their own Multi-Level Intermediate Representation (MLIR) dialect, so porting a circuit to a Quantum Intermediate Representation (QIR) runtime needs a hand-written pass per dialect. Our translator selects among three paths: a deterministic compiler path for known grammars (Catalyst and Quake), a large language model (LLM) agentic path for unseen dialects, and a repair path invoked when deterministic output fails testing. A dual-backend pipeline executes the source MLIR and generated QIR on independent simulators, comparing output distributions by total variation distance (TVD). For smaller open-weight models we replace retrieval-augmented generation with inline context engineering and add a guardrail-assembler that rebuilds valid QIR from incomplete output. Across 41 circuits spanning three dialects and 1--100 qubits, the deterministic path is gate-exact on all 70 circuit-dialect pairs, averaging above 95\% TVD similarity (1-TVD). On an unseen fault-tolerant (FTQC) dialect, four of six models produce valid logical-level QIR on the first iteration.
\end{abstract}


\section{Introduction}

Quantum computing is undergoing rapid diversification in hardware and software~\cite{preskill2018quantum, heim2020quantum}. As vendors and research groups introduce competing frameworks (IBM Qiskit~\cite{javadi2024qiskit}, Google Cirq~\cite{cirq2024}, Microsoft Q\#~\cite{svore2018qsharp}, Xanadu PennyLane~\cite{bergholm2022pennylane}, NVIDIA CUDA-Q~\cite{kim2023cudaquantum}), the software stack has grown fragmented. Each framework defines its own IR for expressing quantum circuits, creating an interoperability barrier: circuits authored in one framework cannot run on another platform's backend without manual or semi-automated translation~\cite{cardama2025review}.

This barrier is sharpest where quantum resources are integrated into scientific HPC ecosystems and the quantum processor is treated as one accelerator among many~\cite{beck2024integrating}. Concretely, a group authoring a variational algorithm in PennyLane Catalyst for its native autodiff may need to dispatch it through a cross-platform QIR runtime such as QIR-EE~\cite{wong2025qiree}, which accepts only QIR; the circuit itself does not change, only its representation, yet translation is currently a hand-written pass rewritten per dialect, the interoperability problem our translator addresses.

MLIR~\cite{lattner2021mlir} was originally developed to address ``end of Moore's law'' challenges in classical compiler infrastructure, and has emerged as a unifying substrate for quantum compilation: PennyLane Catalyst~\cite{ittah2024catalyst} introduces a quantum dialect for just-in-time compilation of hybrid quantum-classical programs, while CUDA-Q defines the Quake dialect for GPU-accelerated quantum simulation~\cite{kim2023cudaquantum}. McCaskey and Nguyen~\cite{mccaskey2021mlirdialect} first demonstrated MLIR dialects' viability for quantum assembly languages, and Hopf et al.~\cite{hopf2026integrating} recently explored integrating multiple quantum tools within the MLIR ecosystem.
\begin{table*}[t]
\centering
\caption{Comparison of approaches to quantum code translation and generation. Our system uniquely combines LLM-based translation with dual-backend simulation testing and support for multiple MLIR dialects.}
\label{tab:comparison}
\renewcommand{\arraystretch}{1.15}
\begin{tabular}{@{}lcccccc@{}}
\toprule
\textbf{System} & \textbf{Source} & \textbf{Target} & \textbf{LLM-Based} & \textbf{Multi-Agent} & \textbf{Sim. Testing} & \textbf{Extensible to New Dialects} \\
\midrule
Catalyst Compiler~\cite{ittah2024catalyst}        & MLIR (Catalyst) & Machine code & No  & No  & No  & No  \\
CUDA-Q~\cite{kim2023cudaquantum}             & MLIR (Quake)    & QIR          & No  & No  & No  & No  \\
Qiskit Code Assistant~\cite{dupuis2024qiskitassistant} & NL / Code   & Qiskit Python & Yes & No  & No  & N/A \\
Agent-Q~\cite{jern2025agentq}                      & NL              & OpenQASM 3   & Yes & No  & Yes & N/A \\
Hopf et al.~\cite{hopf2026integrating}             & MLIR            & MLIR         & No  & No  & No  & Limited \\
\textbf{This work}                                  & MLIR (dialect-agnostic)      & QIR          & Hybrid & Yes & \textbf{Dual-backend} & \textbf{Yes} \\
\bottomrule
\end{tabular}
\end{table*}
QIR~\cite{qirspec2021} is built atop LLVM IR~\cite{lattner2004llvm} and has gained traction as a portable target for heterogeneous backends~\cite{wong2025qiree,qireerepo,QIRRunner}, defining standard quantum intrinsics that many simulators and hardware backends can consume. Bridging MLIR quantum dialects to QIR currently requires hand-written passes per source dialect, an approach that scales poorly as new dialects appear~\cite{cardama2025review, hopf2026integrating}.

Recent advances in LLMs show strong competence in code generation~\cite{chen2021codex, li2022alphacode} and translation~\cite{szafraniec2023code}, including quantum-specific efforts~\cite{dupuis2024qiskitassistant,jern2025agentq,afrose2025learning}. Multi-agent orchestration~\cite{hong2024crewai, wu2023autogen} and reasoning paradigms such as ReAct~\cite{yao2023react} and Reflexion~\cite{shinn2023reflexion} extend single-pass generation to multi-step transformation via iterative feedback; Section~\ref{sec:related} surveys this line in depth.

However, applying LLMs to quantum IR translation presents challenges distinct from general-purpose code translation. Quantum IR has strict syntactic requirements (LLVM type annotations, declaration ordering), subtle semantic constraints (qubit pointer indexing, measurement recording protocols), and zero tolerance for hallucinated operations that alter the circuit's quantum behavior. In practice, smaller open-weight models (8B--13B parameters) practical for local deployment frequently produce structurally incomplete output, omitting scaffolding required for valid QIR while preserving the correct gate sequence.

In this paper, we present an agentic translation system that addresses these challenges through a hybrid architecture combining deterministic parsing with LLM-based translation and iterative repair. Our principal contributions are:

\begin{enumerate}
    \item A \textbf{three-path translation strategy} that dynamically selects between deterministic, fully agentic, and hybrid repair modes based on dialect recognition and testing outcomes.
    \item A \textbf{dual-backend testing pipeline} that independently executes the source MLIR and generated QIR on separate quantum simulators (PennyLane Catalyst, CUDA-Q, or other backends via an extensible registry) and QIR Runner, comparing output probability distributions via total variation distance (TVD).
    \item A \textbf{guardrail-assembler pattern} that recovers valid QIR from structurally incomplete LLM output by extracting gate operations and deterministically reassembling them using validated templates.
    \item A \textbf{context engineering} approach that injects gate mappings and QIR templates directly into agent prompts, eliminating the failure modes associated with retrieval-augmented generation (RAG) on smaller models.
\end{enumerate}

\section{Related Work}\label{sec:related}

We organize the related work into five areas: quantum IRs, MLIR-based quantum compilation, LLM-driven code generation and translation, agentic AI for software engineering, and verification of quantum circuit transformations.

\subsection{Quantum Intermediate Representations}

The proliferation of quantum hardware platforms has motivated multiple intermediate representations as abstraction layers between high-level quantum programs and low-level control pulses. OpenQASM~3~\cite{cross2022openqasm3} extends OpenQASM with classical control flow, subroutines, and timing primitives as a widely adopted circuit-level interchange format. The QIR specification~\cite{qirspec2021} takes a complementary approach, encoding quantum operations as intrinsic function calls within LLVM IR~\cite{lattner2004llvm}, inheriting LLVM's optimization infrastructure and classical-pipeline interoperability.

Cardama et al.~\cite{cardama2025review} survey quantum IRs, comparing QIR, QSSA, and MLIR-based dialects on expressiveness, optimization support, and tooling maturity. Ittah et al.~\cite{ittah2022qiro} proposed QIRO, an SSA-based MLIR program representation supporting dataflow-driven optimizations. Luo and Zhao~\cite{luo2023formalization} formalized quantum IR type-safety, showing well-typed programs preserve qubit linearity and measurement consistency. None of these address automated cross-dialect translation.

\subsection{MLIR for Quantum Compilation}

MLIR~\cite{lattner2021mlir} for quantum computing was pioneered by McCaskey and Nguyen~\cite{mccaskey2021mlirdialect}, building on XACC~\cite{mccaskey2020xacc}, who showed that its lowering and dialect system suit quantum operations at multiple abstraction levels. PennyLane Catalyst~\cite{ittah2024catalyst} realizes this with a quantum dialect (\texttt{quantum.alloc}, \texttt{quantum.custom}, \texttt{quantum.measure}) for JIT compilation of hybrid quantum-classical programs; CUDA-Q~\cite{kim2023cudaquantum} defines the similar Quake dialect, lowering to QIR. Hopf et al.~\cite{hopf2026integrating} integrated Catalyst with the Munich Quantum Toolkit via MLIR; this advances intra-MLIR interoperability but relies on manually authored passes per dialect pair, the limitation motivating our agentic approach.

\subsection{LLM-Driven Code Generation and Translation}

Large language models have achieved strong performance on code generation benchmarks~\cite{chen2021codex, li2022alphacode, austin2021program}, with models such as GPT-4~\cite{achiam2023gpt4} and Code Llama~\cite{roziere2023codellama} producing functionally correct programs from natural-language specifications. Szafraniec et al.~\cite{szafraniec2023code} showed that compiler IRs (LLVM IR) as a pivot improve neural code translation, directly relevant to our MLIR-to-QIR task.

In the quantum domain, Dupuis et al.~\cite{dupuis2024qiskitassistant} developed the Qiskit Code Assistant by fine-tuning Granite LLMs on quantum code, and Jern et al.~\cite{jern2025agentq} proposed Agent-Q for quantum circuit generation and optimization in OpenQASM~3. Both target high-level SDKs (Qiskit Python, OpenQASM) rather than low-level LLVM-based representations like QIR, which impose stricter syntactic requirements. A complementary line of work applies machine learning to classical compiler optimization: Cummins et al.~\cite{cummins2020programl} introduced ProGraML, a graph-based framework for program optimization over LLVM IR. We differ in using LLMs for cross-IR \emph{translation} rather than intra-IR optimization, with correctness defined by quantum-mechanical equivalence of output distributions.

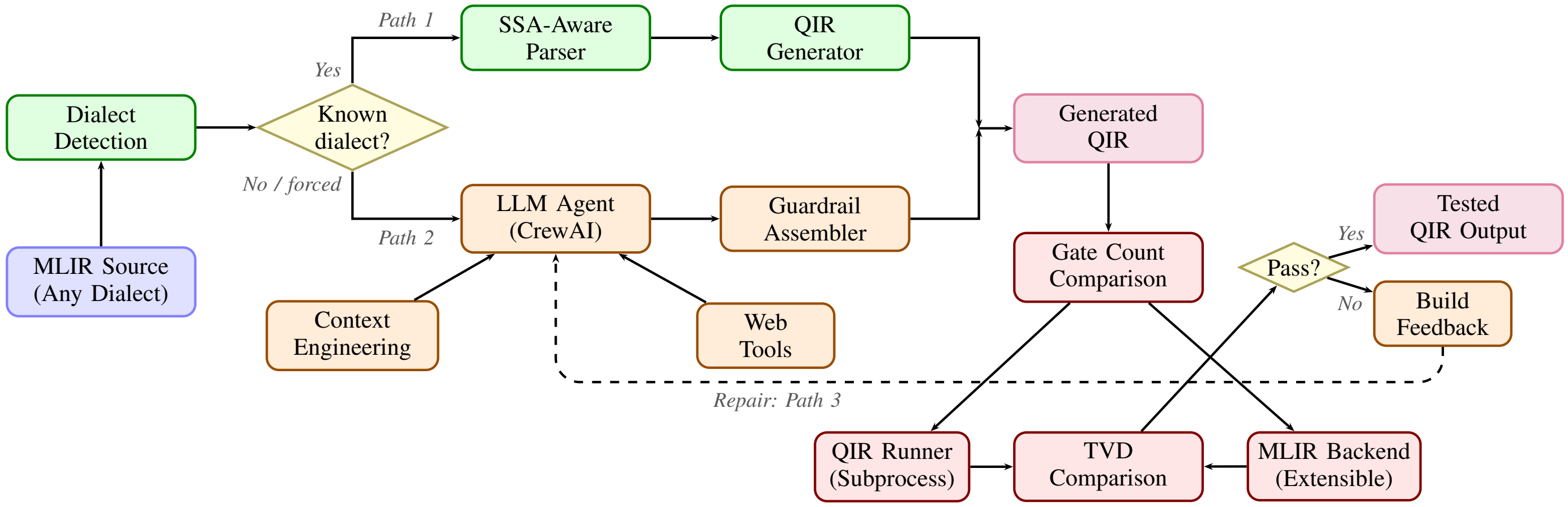
\begin{figure*}[t]
\centering
\begin{tikzpicture}[
    node distance=0.42cm and 0.6cm,
    box/.style={draw, rounded corners, minimum height=0.75cm, minimum width=2.2cm, align=center, font=\footnotesize},
    input/.style={box, fill=blue!12, draw=blue!50, thick},
    det/.style={box, fill=green!12, draw=green!50!black, thick},
    agent/.style={box, fill=orange!15, draw=orange!60!black, thick},
    verify/.style={box, fill=red!10, draw=red!50!black, thick},
    output/.style={box, fill=purple!12, draw=purple!50, thick},
    decision/.style={diamond, draw, thick, fill=yellow!15, draw=yellow!60!black, aspect=2.2, inner sep=1pt, font=\footnotesize, align=center},
    arrow/.style={-{Stealth[length=3pt]}, thick},
    dasharrow/.style={-{Stealth[length=3pt]}, thick, dashed},
    label/.style={font=\scriptsize\itshape, text=gray!70!black},
]

\node[input] (mlir) {MLIR Source\\(Any Dialect)};

\node[det, above=1.0cm of mlir] (detect) {Dialect\\Detection};

\node[decision, right=0.7cm of detect] (known) {Known\\dialect?};

\node[det, above right=0.4cm and 0.7cm of known] (parser) {SSA-Aware\\Parser};
\node[det, right=0.8cm of parser] (qirgen) {QIR\\Generator};

\node[agent, below right=0.4cm and 0.7cm of known] (llmagent) {LLM Agent\\(CrewAI)};

\node[agent, below=1.5cm of known, minimum width=2.0cm] (context) {Context\\Engineering};
\node[agent, right=3cm of context, minimum width=1.6cm] (webtool) {Web\\Tools};

\node[agent, right=0.8cm of llmagent] (guardrail) {Guardrail\\Assembler};

\node[coordinate, right=0.8cm of qirgen] (merge1) {};
\node[coordinate, right=0.8cm of guardrail] (merge2) {};
\node[coordinate] at ($(merge1)!0.5!(merge2)$) (mergepoint) {};
\node[output, right=0.4cm of mergepoint] (qir) {Generated\\QIR};

\node[verify, below=0.8cm of qir] (gatecomp) {Gate Count\\Comparison};

\node[verify, below left=1.5cm and 0.5cm of gatecomp, minimum width=1.8cm] (qirrun) {QIR Runner\\(Subprocess)};
\node[verify, below right=1.5cm and 0.5cm of gatecomp, minimum width=1.8cm] (catrun) {MLIR Backend\\(Extensible)};

\node[verify, below=1.5cm of gatecomp] (tvd) {TVD\\Comparison};

\node[decision, right=0.4cm of gatecomp] (pass) {Pass?};

\node[output, above right=0.0cm and 0.6cm of pass] (success) {Tested\\QIR Output};
\node[agent, below right=0.0cm and 0.6cm of pass, minimum width=1.6cm] (feedback) {Build\\Feedback};

\draw[arrow] (mlir) -- (detect);
\draw[arrow] (detect) -- (known);

\draw[arrow] (known) |- node[label, left, pos=0.15] {Yes} node[label, above, pos=0.75] {Path 1} (parser);
\draw[arrow] (parser) -- (qirgen);
\draw[arrow] (qirgen) -- (merge1) -- (mergepoint);

\draw[arrow] (known) |- node[label, left, pos=0.15] {No / forced} node[label, below, pos=0.75] {Path 2} (llmagent);
\draw[arrow] (llmagent) -- (guardrail);
\draw[arrow] (guardrail) -- (merge2) -- (mergepoint);

\draw[arrow] (context) -- (llmagent);
\draw[arrow] (webtool) -- (llmagent);

\draw[arrow] (mergepoint) -- (qir);
\draw[arrow] (qir) -- (gatecomp);

\draw[arrow] (gatecomp) -- (qirrun);
\draw[arrow] (gatecomp) -- (catrun);

\draw[arrow] (qirrun) -- (tvd);
\draw[arrow] (catrun) -- (tvd);

\draw[arrow] (tvd) -- (pass);


\draw[arrow] (pass) -- node[label, above] {Yes} (success);

\draw[arrow] (pass) -- node[label, below] {No} (feedback);
\draw[dasharrow, rounded corners=8pt] (feedback) -- ++(0,-0.8) coordinate (rp3bend) -| (llmagent.south);
\node[label, below] at ($(llmagent.south |- rp3bend)!0.25!(rp3bend)$) {Repair: Path 3};

\end{tikzpicture}
\caption{System architecture. \textbf{Path~1} (deterministic, top): recognized dialects. \textbf{Path~2} (agentic, bottom): unseen dialects or forced mode. \textbf{Path~3} (repair, dashed): deterministic output that fails testing is repaired by the LLM agent. The testing pipeline first compares gate counts, then executes QIR and MLIR independently on separate backends and compares distributions via TVD. The ``Pass?'' decision requires both gate match and $1{-}\text{TVD} \geq 95\%$.}
\label{fig:architecture}
\end{figure*}

\subsection{Agentic AI for Software Engineering}

LLM-based agents combining reasoning with tool use were formalized by ReAct~\cite{yao2023react}, interleaving chain-of-thought reasoning~\cite{wei2022chainofthought} with actions like code execution and search. Shinn et al.~\cite{shinn2023reflexion} extended this with Reflexion, enabling agents to improve via verbal self-reflection on execution feedback, closely related to our testing-feedback loop.

Multi-agent frameworks such as AutoGen~\cite{wu2023autogen} and CrewAI~\cite{hong2024crewai} orchestrate specialized agents on complex tasks; Wang et al.~\cite{wang2024agentsurvey} survey LLM-based agents for software engineering. We extend these to quantum compiler engineering: a translation agent generates QIR and a testing pipeline provides structured feedback for repair, with context engineering over RAG~\cite{lewis2020rag} adapting to failure modes observed with smaller models.

\subsection{Quantum Circuit Verification and Testing}

Verifying the correctness of quantum circuit transformations is a recognized challenge in quantum compilation~\cite{karuppasamy2025comprehensive}. Bauer-Marquart et al.~\cite{bauer2023symqv} proposed symQV, a symbolic verification approach that proves equivalence of quantum programs through automated theorem proving. While formally rigorous, symbolic approaches face scalability limitations for circuits with many qubits or complex classical control flow.

Simulation-based testing offers a complementary approach: executing both circuits on simulators and comparing output distributions detects semantic discrepancies without formal proofs. Total variation distance (TVD)~\cite{nielsen2010quantum} provides a symmetric, bounded measure of distributional agreement. Our dual-backend pipeline implements this strategy with the added rigor of executing source MLIR and generated QIR on \emph{independent} backends. We deliberately favor execution‑based \emph{testing} over formal \emph{verification}: our pipeline provides statistical confidence within a threshold, whereas approaches such as symQV offer formal guarantees.

Table~\ref{tab:comparison} positions our work relative to existing quantum translation and code generation approaches.

\section{Methodology}\label{sec:methodology}

This section presents the architecture and algorithms of the agentic MLIR-to-QIR translation system. The system consists of four principal components: dialect-aware parsing, multi-path translation, dual-backend testing, and an LLM output guardrail with deterministic reassembly. Fig.~\ref{fig:architecture} provides an overview of the system architecture and the interaction between components.

\subsection{Dialect Detection and SSA-Aware Parsing}\label{sec:parsing}

The translation pipeline begins with automatic dialect detection. The system scans the input MLIR source for dialect-specific markers: the presence of \texttt{quantum.custom}, \texttt{catalyst.launch\_kernel}, or \texttt{quantum.alloc} tokens identifies the Catalyst dialect~\cite{ittah2024catalyst}, while \texttt{quake.alloca}, \texttt{!quake.veq}, or \texttt{quake.h} tokens identify the Quake dialect~\cite{kim2023cudaquantum}. If no known markers are found, the dialect is classified as unseen, triggering the agentic translation path.

For recognized dialects, the system employs an SSA-aware parser that traverses the MLIR program in a single pass while maintaining three dictionaries that track the dataflow through the circuit. The first dictionary maps MLIR variable names to physical qubit indices, resolving \texttt{quantum.extract} operations that bind SSA values to specific qubits in the allocated register. The second dictionary records constant values declared by \texttt{arith.constant} operations, which supply rotation angles for parameterized gates such as RX, RY, and RZ. The third dictionary traces measurement results through the SSA chain, enabling the parser to correctly associate conditional operations (\texttt{scf.if} blocks) with their controlling measurement outcomes.

Multi-result gates are the main parsing challenge: in Catalyst a CNOT produces two output SSA values, one per qubit, and the parser resolves both to physical qubit indices by tracking operand order. The parser emits an ordered operation list that preserves the original gate sequence, measurement positions, and conditional block boundaries, producing an \texttt{MLIRCircuit} data structure consumed by both the deterministic QIR generator and the testing pipeline.

\subsection{Three-Path Translation Strategy}\label{sec:translation}

The system implements three translation paths, selected dynamically based on dialect recognition and testing outcomes. Algorithm~\ref{alg:translation} summarizes the routing logic: dialect recognition picks Path~1 or Path~2 as mutually exclusive entry points, and Path~1 falls through to Path~3 on a testing failure, so a translation takes exactly one of Path~1, Path~2, or Path~1-then-3.

\begin{algorithm}[h]
\caption{Translation Path Selection and Testing Loop}
\label{alg:translation}
\begin{algorithmic}[1]
\REQUIRE MLIR source code $M$, maximum iterations $K$, similarity threshold $\tau$
\ENSURE QIR code $Q$, testing result $V$
\STATE $d \leftarrow \textsc{DetectDialect}(M)$
\STATE $Q \leftarrow \emptyset$, $V.\text{similarity} \leftarrow 0$ \COMMENT{sentinel for Path 2}

\IF{$d \in \{\text{catalyst}, \text{quake}\}$ \AND \NOT force\_agentic}
    \STATE $Q \leftarrow \textsc{DeterministicTranslate}(M, d)$ \COMMENT{Path 1}
    \STATE $V \leftarrow \textsc{Test}(M, Q)$
    \IF{$V.\text{passed}$}
        \RETURN $Q, V$ \COMMENT{Deterministic success}
    \ENDIF
    \STATE \textbf{go to} \textsc{AgenticLoop} \COMMENT{Path 3: repair with agentic}
\ELSE
    \STATE \textbf{go to} \textsc{AgenticLoop} \COMMENT{Path 2: agentic}
\ENDIF
\STATE \textsc{AgenticLoop}:
\STATE $Q_{\text{best}} \leftarrow Q$, $s_{\text{best}} \leftarrow V.\text{similarity}$, $V_{\text{best}} \leftarrow V$
\FOR{$i = 1$ \TO $K$}
    \STATE $F \leftarrow \textsc{BuildFeedback}(V)$
    \STATE $Q \leftarrow \textsc{AgentTranslate}(M, d, F, Q_{\text{best}}, i)$
    \STATE $Q \leftarrow \textsc{Guardrail}(Q)$ \COMMENT{Reassemble if needed}
    \STATE $V \leftarrow \textsc{Test}(M, Q)$
    \IF{$V.\text{gate\_match}$ \AND $V.\text{similarity} \geq \tau$}
        \RETURN $Q, V$
    \ENDIF
    \IF{$V.\text{similarity} > s_{\text{best}}$}
        \STATE $Q_{\text{best}} \leftarrow Q$ \\ 
        \STATE $s_{\text{best}} \leftarrow V.\text{similarity}$ 
        \\ \STATE 
        $V_{\text{best}} \leftarrow V$
    \ENDIF
\ENDFOR
\RETURN $Q_{\text{best}}, V_{\text{best}}$
\end{algorithmic}
\end{algorithm}

\textbf{Path 1: Deterministic translation.} When the detected dialect is recognized, the parser extracts the circuit structure and the QIR generator emits complete LLVM IR by instantiating validated templates. The generator maps each MLIR gate name to its corresponding QIR intrinsic (e.g., \texttt{quantum.custom "Hadamard"} to \texttt{@\_\_quantum\_\_qis\_\_h\_\_body}), computes qubit pointer expressions using the QIR addressing convention (null for qubit 0, \texttt{inttoptr (i64 N to \%Qubit*)} for qubit $N$), and appends measurement operations, output recording calls, function declarations, and QIR metadata. This path requires no LLM invocation and completes in 1--26~ms across the benchmark suite (Section~\ref{sec:results}).

\textbf{Path 2: Fully agentic translation.} When the dialect is unseen or the user explicitly forces agentic mode, the system delegates translation to an LLM agent orchestrated through the CrewAI framework~\cite{hong2024crewai}. The agent receives the MLIR source along with an inline reference containing gate mapping tables, a complete QIR example, and translation pattern descriptions. This context engineering approach injects all necessary domain knowledge directly into the prompt rather than relying on retrieval-augmented generation (RAG)~\cite{lewis2020rag}, which we found unreliable with smaller models that frequently misuse retrieval tools. The agent is instructed to produce output in one of two formats: raw LLVM IR (preferred) or a JSON object containing an operations list with qubit indices, which the guardrail can then assemble into valid QIR (Section~\ref{sec:guardrail}). When the dialect is unseen, the agent also has access to a web search tool for fetching documentation about unfamiliar MLIR namespace prefixes.

A \texttt{--force-agentic} flag routes a \emph{recognized} dialect through the agentic path anyway, letting a user benchmark agentic output against known-correct ground truth (Section~\ref{sec:results}); it is not part of Algorithm~\ref{alg:translation}'s automatic routing.

\textbf{Path 3: Deterministic with agentic repair.} When the deterministic path produces QIR that fails testing (Section~\ref{sec:testing}), the system transitions to an iterative repair loop. The testing failure is converted into structured feedback that specifies which gates are missing or extraneous, whether the output distribution diverges from the expected reference, and the measured TVD similarity score. This feedback, together with the previous QIR attempt, is provided to the LLM agent, which generates a corrected translation. The loop continues for up to $K$ iterations (configurable, default $K=5$), tracking the highest-scoring attempt to ensure the system returns its best result even if the maximum iteration count is reached without full convergence. This iterative refinement pattern is inspired by the Reflexion framework~\cite{shinn2023reflexion}, adapted from natural language tasks to quantum IR translation.

Table~\ref{tab:gate_mapping} shows the gate mapping used for both deterministic translation and context engineering in the agentic path.

\begin{table}[htbp]
\centering
\caption{MLIR gate name to QIR intrinsic mapping. Single-qubit gates map to unary QIR functions; two-qubit gates map to binary functions with control and target qubit arguments.}
\label{tab:gate_mapping}
\renewcommand{\arraystretch}{1.1}
\begin{tabular}{@{}lll@{}}
\toprule
\textbf{MLIR Gate} & \textbf{QIR Intrinsic Suffix} & \textbf{Params} \\
\midrule
Hadamard       & \texttt{@\_\_quantum\_\_qis\_\_h\_\_body}       & --- \\
PauliX         & \texttt{@\_\_quantum\_\_qis\_\_x\_\_body}       & --- \\
PauliY         & \texttt{@\_\_quantum\_\_qis\_\_y\_\_body}       & --- \\
PauliZ         & \texttt{@\_\_quantum\_\_qis\_\_z\_\_body}       & --- \\
S              & \texttt{@\_\_quantum\_\_qis\_\_s\_\_body}       & --- \\
T              & \texttt{@\_\_quantum\_\_qis\_\_t\_\_body}       & --- \\
RX             & \texttt{@\_\_quantum\_\_qis\_\_rx\_\_body}      & $\theta$ \\
RY             & \texttt{@\_\_quantum\_\_qis\_\_ry\_\_body}      & $\theta$ \\
RZ             & \texttt{@\_\_quantum\_\_qis\_\_rz\_\_body}      & $\theta$ \\
CNOT           & \texttt{@\_\_quantum\_\_qis\_\_cnot\_\_body}    & --- \\
CZ             & \texttt{@\_\_quantum\_\_qis\_\_cz\_\_body}      & --- \\
SWAP           & \texttt{@\_\_quantum\_\_qis\_\_swap\_\_body}    & --- \\
\bottomrule
\end{tabular}
\end{table}

\subsection{Dual-Backend Testing Pipeline}\label{sec:testing}

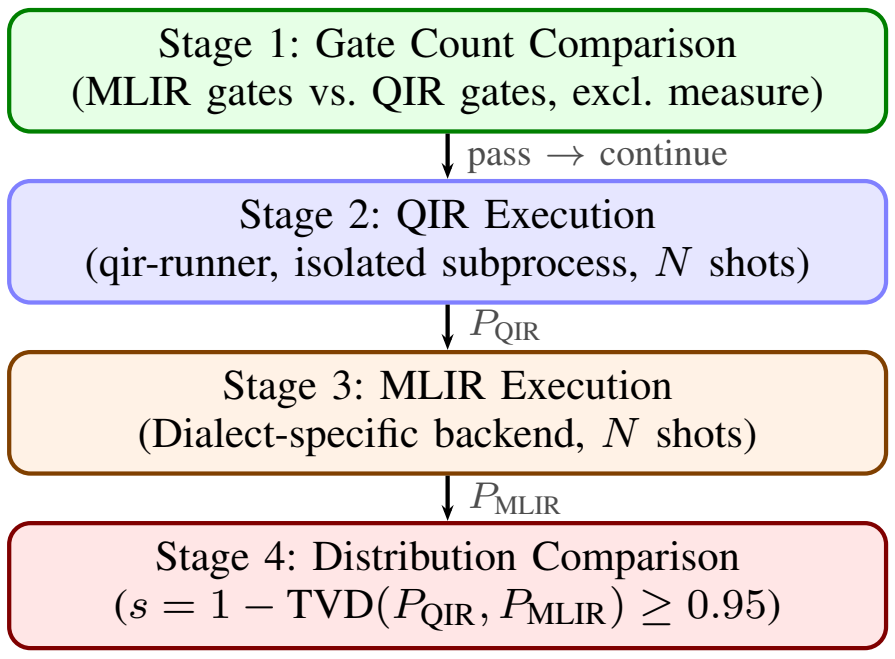
\begin{figure}[htbp]
\centering
\begin{tikzpicture}[
    node distance=0.3cm,
    stage/.style={draw, rounded corners, minimum height=0.55cm, minimum width=5.8cm, align=center, font=\footnotesize, thick},
    s1/.style={stage, fill=green!10, draw=green!50!black},
    s2/.style={stage, fill=blue!10, draw=blue!50},
    s3/.style={stage, fill=orange!10, draw=orange!50!black},
    s4/.style={stage, fill=red!10, draw=red!50!black},
    arrow/.style={-{Stealth[length=3pt]}, thick},
    lbl/.style={font=\scriptsize, text=gray!60!black},
]
\node[s1] (s1) {Stage 1: Gate Count Comparison\\(MLIR gates vs. QIR gates, excl. measure)};
\node[s2, below=of s1] (s2) {Stage 2: QIR Execution\\(qir-runner, isolated subprocess, $N$ shots)};
\node[s3, below=of s2] (s3) {Stage 3: MLIR Execution\\(Dialect-specific backend, $N$ shots)};
\node[s4, below=of s3] (s4) {Stage 4: Distribution Comparison\\($s = 1 - \text{TVD}(P_\text{QIR}, P_\text{MLIR}) \geq 0.95$)};

\draw[arrow] (s1) -- node[lbl, right] {pass $\rightarrow$ continue} (s2);
\draw[arrow] (s2) -- node[lbl, right] {$P_\text{QIR}$} (s3);
\draw[arrow] (s3) -- node[lbl, right] {$P_\text{MLIR}$} (s4);
\end{tikzpicture}
\caption{Four-stage dual-backend testing pipeline. Each stage provides progressively deeper correctness assurance, from structural gate comparison to distributional equivalence via TVD.}
\label{fig:testing}
\end{figure}

A central contribution of this work is the dual-backend testing pipeline, which provides an execution-based correctness check that goes beyond syntactic or structural validation. The pipeline operates in four stages, each progressively more computationally expensive. Fig.~\ref{fig:testing} illustrates the testing flow.

\textbf{Stage 1: Gate count comparison.} The system independently counts gates in the original MLIR and the generated QIR. For MLIR, gate names are extracted from \texttt{quantum.custom} (Catalyst) or dialect-specific operations (Quake) and normalized to lowercase canonical names. For QIR, gate names are extracted from \texttt{call void @\_\_quantum\_\_qis\_\_<gate>\_\_body} patterns. Measurement operations are excluded from both counts because they are non-unitary, and testing appends terminal measurements on all qubits (Stage 2) that would otherwise inflate the QIR count. A gate count mismatch provides an immediate signal that the translation introduced or dropped operations.

\textbf{Stage 2: QIR execution.} The generated QIR is executed using the \texttt{qir-runner} package (version 0.9.1)~\cite{QIRRunner}, which provides a Python interface to a QIR execution engine. To avoid conflicts between the LLVM libraries linked by \texttt{qir-runner} and the JAX/XLA runtime used by PennyLane Catalyst, execution is performed in an isolated subprocess. Before execution, a preprocessing step ensures that measurement operations cover all qubits in the circuit and that output recording calls are present in the correct order. The executor runs the circuit for $N$ shots (default $N=1024$) and parses the output into a bitstring frequency distribution $P_{\text{QIR}}$. Note that in this stage, terminal measurements are appended on all qubits to normalize distribution comparison. Circuits with mid‑circuit measurements rely on dialect‑specific reconstruction of classical feed‑forward in Stage 3.

\textbf{Stage 3: MLIR execution.} The original MLIR circuit is independently executed on a dialect-specific backend selected through an extensible simulator registry. For the Catalyst dialect, the system reconstructs a PennyLane quantum node (\texttt{QNode}) from the parsed circuit, applies the \texttt{@catalyst.qjit} just-in-time compiler with the \texttt{lightning.qubit} simulator~\cite{bergholm2022pennylane}, and collects measurement samples. Circuits with mid-circuit measurements use a snapshot-based construction that replays gate sequences inside the JIT-compiled function via Catalyst's \texttt{@catalyst.cond} decorator. For the Quake dialect, the backend is CUDA-Q~\cite{kim2023cudaquantum}. Both backends execute in isolated subprocesses to avoid LLVM library conflicts, producing a second bitstring frequency distribution $P_{\text{MLIR}}$. The registry admits additional backends for new dialects; where none is yet available, no $P_{\text{MLIR}}$ can be produced and the pipeline falls back to QIR-only testing, confirming the QIR is executable and reporting QIR-side gate counts for structural inspection. The same interface is the attachment point for hardware and HPC-hosted backends, although every backend registered in this work is a local simulator.

\textbf{Stage 4: Distribution comparison.} The two output distributions are compared using the total variation distance (TVD), defined as:

\begin{equation}
\text{TVD}(P, Q) = \frac{1}{2} \sum_{x \in \mathcal{X}} |P(x) - Q(x)|
\label{eq:tvd}
\end{equation}

\noindent where $\mathcal{X}$ is the set of all observed bitstring outcomes across both distributions, and $P(x)$ and $Q(x)$ are the empirical probabilities of outcome $x$ under each distribution. The TVD ranges from 0 (identical distributions) to 1 (completely disjoint support). We convert TVD to a similarity score $s = 1 - \text{TVD}$ and apply a pass threshold of $\tau = 0.95$, meaning the translation is accepted when the output distributions agree to within 5\% total variation. This threshold accounts for finite sampling noise inherent in shot-based simulation while remaining sensitive to gate errors that would produce qualitatively different output states.

We choose TVD over Kullback-Leibler (KL) divergence for two reasons. First, TVD is symmetric, which avoids the arbitrary choice of which distribution serves as the reference. Second, TVD is well-defined even when one distribution assigns zero probability to an outcome observed in the other, whereas KL divergence diverges in this case unless smoothing is applied.

\begin{table*}[t]
\centering
\caption{Benchmark circuit suite: 41 unique circuit designs, 73 circuit-dialect instances. Circuits are organized by category with qubit counts, gate types, dialect availability, and instance count per row}
\label{tab:circuits}
\renewcommand{\arraystretch}{1.1}
\begin{tabular}{@{}llclll@{}}
\toprule
\textbf{Category} & \textbf{Circuit} & \textbf{Qubits} & \textbf{Key Gates} & \textbf{Dialects} & \textbf{Count} \\
\midrule
Standard                     & Bell state             & 2   & H, CNOT       & Both (Catalyst \& Quake) & 2 \\
\midrule
GHZ scaling & GHZ-\{5, 10, 15, ..., 100\} & 5--100 & H, CNOT$\times(N{-}1)$ & Both & 40 \\
\midrule
\multirow{2}{*}{MBQC}       & Teleport, RZ rotation  & 2   & H, CZ, X, conditional  & Both & 4 \\
                             & RX rotation, CNOT      & 3--4 & H, CZ, X, Z, conditional & Both & 4 \\
\midrule
Conditional                  & Classic teleportation  & 3   & H, CNOT, X, Z & Both & 2 \\
\midrule
Random                       & 5 mixed-gate circuits  & 2--7 & T, S, RX, RY, RZ, CZ, SWAP & Both & 10 \\
\midrule
QEC                          & Steane code on 1 logical qubit           & 7   & H$\times$10, CNOT$\times$9 & Both & 2 \\
\midrule
Parametric                   & RX/RY/RZ rotations     & 1   & RX, RY, RZ    & Quake only & 3 \\
\midrule
Variational                  & Autodiff gradient      & 1   & RX, ExpVal(PauliZ)    & Catalyst only & 1 \\
\midrule
\multirow{2}{*}{Algorithms}  & Grover (3Q, 2/8 marked) & 3   & H, X, T, RY, RZ, CNOT & Quake only & 1 \\
                             & Shor order-finding ($N{=}15$) & 12  & H, X, T, RZ, CNOT & Quake only & 1 \\
\midrule
\multirow{2}{*}{FTQC (unseen)} & Steane 1Q H, 2Q Bell & 1--2\textsuperscript{$\dagger$} & Logical H, CNOT & FTQC only & 2 \\
                             & Steane 3Q Grover & 3\textsuperscript{$\dagger$}   & Logical H, CZ, Z & FTQC only & 1 \\
\midrule
\multicolumn{5}{@{}r}{\textbf{Total circuit-dialect instances}} & \textbf{73} \\
\bottomrule
\multicolumn{6}{@{}l}{\footnotesize \textsuperscript{$\dagger$}Logical qubits; physical count is 7-times the number of qubits under Steane [[7,1,3]] encoding.} 
\end{tabular}
\end{table*}

\subsection{Guardrail-Assembler Pattern for LLM Output Recovery}\label{sec:guardrail}

Smaller LLMs (8B to 13B) often produce correct gate sequences but incomplete LLVM/QIR scaffolding, omitting type definitions (\texttt{\%Qubit = type opaque}), declarations, attributes, or metadata. Some emit a JSON operations list rather than raw LLVM IR, or tool-call JSON when confused by the agent framework's tool-use protocol.

To recover valid QIR from these partial outputs, we introduce a guardrail-assembler pattern that operates as a post-processing stage after each LLM invocation. The guardrail applies the following cascade of five checks, in order of priority, each falling through to the next on failure: (1) is the output already complete QIR? pass through; (2) are gate calls present? extract and assemble; (3) is it a JSON operations list? parse and assemble; (4) is it wrapped in markdown or prose? strip and retry; (5) otherwise, unrecoverable: return error feedback to the agent.

The deterministic QIR assembler accepts a list of parsed gate operations and produces syntactically complete LLVM IR by instantiating the same templates used by the deterministic translation path. It normalizes gate names through an alias table that maps common LLM variants (e.g., ``hadamard'', ``cx'', ``pauli\_x'') to canonical QIR function suffixes, infers the qubit count from the maximum qubit index referenced in the operations, and appends measurement operations for all qubits, output recording calls, function declarations, attributes, and QIR version metadata.

\subsection{Context Engineering for LLM Translation}\label{sec:context}

Domain knowledge is usually supplied to LLM agents through retrieval-augmented generation (RAG)~\cite{lewis2020rag}. In preliminary experiments, 8B models frequently misused the retrieval tool, issuing repeated identical queries or misinterpreting retrieved content, so we inject all reference material directly into the task prompt.

The inline reference consists of three components: (1) a gate mapping table that associates each MLIR gate name with its QIR intrinsic function signature (Table~\ref{tab:gate_mapping}), covering single-qubit gates, parameterized rotation gates, and two-qubit gates; (2) a complete QIR example showing a two-qubit Bell state circuit with all required sections (header, entry function, declarations, attributes, and metadata); and (3) a set of translation pattern descriptions covering qubit allocation, gate sequencing, measurement recording, and conditional operations.

The added prompt length is modest: the gate mapping table and QIR template total roughly 2,000 tokens, acceptable even for 4,096-token context windows. For unseen dialects the agent also retains a web search tool for external documentation.

\subsection{Multi-Model Support}\label{sec:models}

The system supports multiple LLM backends through a unified configuration layer: local models via Ollama (Llama 3.1~\cite{dubey2024llama3} 8B/70B, Code Llama~\cite{roziere2023codellama} 13B/34B) and cloud models, GPT-OSS-20B and Gemma~4 (31B), via the HuggingFace Inference API. CrewAI~\cite{hong2024crewai} wraps the selected model in a unified interface for prompt formatting and ReAct~\cite{yao2023react} reasoning.

\textbf{Sampling and model selection.} All models use temperature $0.1$, a 4,096-token limit, provider-default \texttt{top\_p}/\texttt{top\_k}, no hyperparameter sweep, and no few-shot ablation (Section~\ref{sec:discussion}). Models span Llama and Code Llama at multiple scales plus two cloud models, GPT-OSS-20B and Gemma~4; this selection is not exhaustive.

\section{Experimental Setup}\label{sec:experiments}

\subsection{Benchmark Circuits}

We assembled a benchmark suite of 41 unique circuit designs (73 circuit-dialect instances, since Catalyst/Quake designs are implemented in both dialects) spanning ten categories (Table~\ref{tab:circuits}). GHZ-state circuits provide a scaling ladder from 5 to 100 qubits across 20 sizes (every 5 qubits) with identical gate structure (one Hadamard, $N{-}1$ CNOTs). Measurement-based quantum computing (MBQC)~\cite{raussendorf2001oneway} circuits test mid-circuit measurements and classical feed-forward. Random circuits stress-test gate diversity (RX, RY, RZ, S, T, CZ, SWAP). The Steane code circuit is a 7-qubit encoder for the [[7,1,3]] code, using 10 Hadamards and 9 CNOTs. Three circuits in an unseen dialect exercise the agentic path on Steane-encoded logical operations.

We additionally translated Grover's and Shor's algorithms from IBM Quantum's Qiskit tutorials~\cite{ibmgrover2025,ibmshor2025}: each circuit is exported to OpenQASM~2.0, converted to Quake MLIR via an existing CUDA-Q-based frontend, and translated by the same deterministic pass (Quake only; Catalyst needs a Qiskit-to-PennyLane step). Grover marks 2 of 8 3-qubit states (53 gates); Shor performs order-finding for $N{=}15$, $a{=}2$ via the standard compiled SWAP-based technique (390 gates).

For all 20 GHZ sizes, we generated independent QIR ground-truth files via \texttt{qiskit-qir} from equivalent Qiskit \texttt{QuantumCircuit} objects, a third-party structural reference for gate-level comparison. The FTQC circuits use a custom \texttt{ftqc} dialect with Steane-encoded logical qubit operations (\texttt{ftqc.init\_zero}, \texttt{ftqc.logical\_h}, \texttt{ftqc.logical\_cnot}), unregistered in the deterministic parser, exercising the agentic path on a genuinely unseen dialect.

\subsection{Models Under Test}

We evaluated six LLM models spanning three size classes and three architectures (Table~\ref{tab:models}). Local models were served through Ollama; cloud models accessed the HuggingFace Inference API. The deterministic parser (no LLM) served as the baseline for all comparisons.

\begin{table}[htbp]
\centering
\caption{LLM models evaluated. VRAM listed for local Ollama deployment.}
\label{tab:models}
\renewcommand{\arraystretch}{1.0}
\begin{tabular}{@{}llcl@{}}
\toprule
\textbf{Model} & \textbf{Provider} & \textbf{Params} & \textbf{VRAM} \\
\midrule
Llama 3.1      & Ollama (local)     & 8B   & $\sim$6~GB  \\
Llama 3.1      & Ollama (local)     & 70B  & $\sim$40~GB \\
Code Llama     & Ollama (local)     & 13B  & $\sim$8~GB  \\
Code Llama     & Ollama (local)     & 34B  & $\sim$20~GB \\
GPT-OSS-20B    & HuggingFace (API)  & 20B  & Cloud       \\
Gemma 4        & HuggingFace (API)  & 31B  & Cloud       \\
\bottomrule
\end{tabular}
\end{table}
\subsection{Evaluation Metrics}

We report four primary metrics across experiments:

\begin{itemize}
    \item \textbf{Gate match rate}: binary correctness of gate-level translation; MLIR and QIR gate counts (excluding measurement) must match in type and quantity.
    \item \textbf{TVD similarity}: $s = 1 - \text{TVD}(P_\text{QIR}, P_\text{MLIR})$, in \emph{shots mode} (1,024 samples per backend) and \emph{probs mode} (exact MLIR-side probabilities via \texttt{qml.probs()}, 100K shots on QIR side), which isolates translation error from sampling noise.
    \item \textbf{Translation time}: Wall-clock time from MLIR input to QIR output, excluding testing.
    \item \textbf{Iteration count}: Number of agent refinement iterations before convergence (1 for deterministic path).
\end{itemize}

\subsection{Hardware and Software Configuration}\label{sec:hardware}

All experiments ran on a workstation with an NVIDIA RTX 6000 Ada GPU (48~GB VRAM), Intel Xeon w5-2465X, and 256~GB RAM, Ubuntu 22.04. Software: Python~3.12, PennyLane~0.44.0, PennyLane-Catalyst~0.14.0, CUDA-Q~0.13, \texttt{qir-runner}~0.9.1, CrewAI~1.10.0, Ollama~0.6. Each agentic experiment was repeated 3 times to capture LLM output variance. The codebase will be available at \url{https://github.com/ORNL-QCI/Agentic-MLIR_QIR}.

\section{Results}\label{sec:results}

\subsection{Translation Correctness}

Table~\ref{tab:e1} reports deterministic results across both dialects and testing modes: 33 Catalyst and 37 Quake circuits (70 pairs, Quake including Grover and Shor), all correctly translated at the gate level. Circuits $\geq$35 qubits are gate-checked only, excluded from TVD statistics since the $2^N$-dimensional state vector exceeds available memory. Six circuit-dialect instances are further excluded from Simulated/TVD (Table~\ref{tab:e1} caption lists them): the two random circuits in both dialects and Quake's MBQC-CNOT give correct gate counts but near-zero TVD from an MLIR-side mid-circuit-measurement reconstruction gap, while Catalyst's autodiff gradient circuit returns a scalar expectation value and its gradient rather than sampled measurements, leaving no distribution to compare. Quake's Shor probs-mode result is excluded separately, its 100K-shot QIR-side simulation exceeding the \texttt{qir-runner} timeout. Among the remaining simulable circuits, Catalyst TVD averages 96.9\% (shots) / 99.2\% (probs) and Quake averages 97.5\% (shots) / 98.5\% (probs). The shots-vs-probs gap reflects sampling noise: the 91.9\% shots-mode minimum, on Catalyst's Steane circuit, reaches 99.5\% in probs mode where the MLIR-side distribution is exact. Both of these cases should not be viewed as translation errors, but instead as testing backend limitations (see Sec.~\ref{sec:limitations}).

\textbf{Path 3 (repair) trigger rate.} Table~\ref{tab:e1}'s sweep runs the CLI without an LLM configured, so Path~3 is unreachable there regardless of TVD; this is not because every circuit passed. Evaluating the pass decision in probs mode, six circuit-dialect instances fall below $\tau$: the two random circuits in both dialects, Quake's MBQC-CNOT, and Quake's Steane circuit at 94.5\%. Each would trigger Path~3 under Algorithm~\ref{alg:translation}'s full routing. We exercised Path~3 directly on the two known-failing random circuits: all 3 repair attempts (Llama~3.1 8B $\times$2, 70B $\times$1) exhausted $K=5$ iterations without passing. Since the gap traces to an MLIR-side testing limitation (Section~\ref{sec:limitations}), repair here optimizes against a reference distribution that does not itself represent the correct output.

\textbf{Grover and Shor.} Both circuits (Section~\ref{sec:experiments}) are part of the Quake benchmark above and translate with 100\% gate-level correctness (53/53, 390/390 gates). Grover clears 95\% TVD in both modes (99.9\% shots, 98.6\% probs); Shor reaches 93.7\% in shots mode, below threshold, with its probs-mode result excluded (Table~\ref{tab:e1}).

\textbf{Forced agentic vs.\ deterministic ground truth.} Routing the Bell state circuit (translated exactly by Path~1) through Path~2 via \texttt{force\_agentic} is architecture-dependent: Gemma~4 (31B) passed on the first iteration on both Catalyst and Quake (4/4 runs, 97.9--99.6\% TVD), matching Path~1's accuracy at $\sim$35--62~s instead of 1--12~ms. Llama~3.1 (8B and 70B) failed all 3 attempts on the same circuits even after we hardened the guardrail and repair feedback (Section~\ref{sec:guardrail}); 8B still mistranslated Catalyst's non-gate expectation-value ops as fake QIR intrinsics, and 70B never produced valid QIR. The gap is model-dependent rather than a fundamental Path~2 defect; since Llama 3.1 succeeds on FTQC (Section~\ref{result_unseen_dialect}), the failure likely stems from Catalyst-specific constructs such as expectation-value ops rather than parameter count.

\begin{table}[htbp]
\centering
\caption{Deterministic translation correctness by dialect and mode.
Circuits $\geq$35~qubits are gate-checked only. Excluded from
Simulated/TVD: \texttt{random\_1847} and \texttt{random\_2449} (both
dialects), \texttt{mbqc\_cnot} (Quake), \texttt{autodiff\_gradient}
(Catalyst); \texttt{shor\_orderfinding} in Quake Probs only.}

\label{tab:e1}
\renewcommand{\arraystretch}{0.95}
\scriptsize
\renewcommand{\arraystretch}{1.1}
\begin{tabular}{@{}lcccccc@{}}
\toprule
\textbf{Dialect} & \textbf{Mode} & \textbf{Total} & \textbf{Gate OK} & \textbf{Simulated} & \textbf{Avg TVD} & \textbf{Min TVD} \\
\midrule
Catalyst & Shots & 33 & 33 & 16 & 96.9\% & 91.9\% \\
Catalyst & Probs & 33 & 33 & 16 & 99.2\% & 96.3\% \\
Quake    & Shots & 37 & 37 & 20 & 97.5\% & 93.7\% \\
Quake    & Probs & 37 & 37 & 19 & 98.5\% & 94.5\% \\
\bottomrule
\end{tabular}
\end{table}

\subsection{Scalability}

We evaluated GHZ circuits across all 20 sizes (5--100 qubits) in both dialects, 40 pairs (Table~\ref{tab:e2}), each translated once since the deterministic path is fully deterministic. All pairs achieve 100\% gate match; translation time stays between 14--24~ms with no measurable increase from 5 to 100 qubits, so the path is dominated by fixed overhead, not gate count. Among the 12 simulable pairs ($\leq$GHZ-30), shots-mode TVD averages 97.8\% (min 94.8\%) and probs-mode 99.2\% (min 97.9\%). Larger circuits exceed state-vector memory but pass gate checking and match the independently generated Qiskit QIR ground truth.

\begin{table}[htbp]
\centering
\caption{GHZ scalability (40 pairs, 100\% gate match). Left: TVD-simulated sizes ($\leq$30 qubits, averaged across both dialects). Right: larger sizes, gate-checked only, since the $2^N$-dimensional state vector exceeds available memory. Time is shots mode, averaged across dialects, excluding testing.}
\label{tab:e2}
\renewcommand{\arraystretch}{0.95}
\footnotesize
\begin{minipage}[t]{0.52\linewidth}
\centering
\begin{tabular}{@{}rccr@{}}
\toprule
\textbf{Qb} & \textbf{Shots} & \textbf{Probs} & \textbf{ms} \\
\midrule
 5 & 99.0\% & 99.3\% & 20 \\
10 & 97.2\% & 99.7\% & 15 \\
15 & 98.5\% & 99.1\% & 18 \\
20 & 97.6\% & 98.9\% & 18 \\
25 & 97.3\% & 99.3\% & 15 \\
30 & 96.7\% & 99.0\% & 19 \\
\bottomrule
\end{tabular}
\end{minipage}%
\begin{minipage}[t]{0.46\linewidth}
\centering
\begin{tabular}{@{}rr|rr@{}}
\toprule
\textbf{Qb} & \textbf{ms} & \textbf{Qb} & \textbf{ms} \\
\midrule
35 & 14 & 70  & 16 \\
40 & 15 & 75  & 16 \\
45 & 19 & 80  & 16 \\
50 & 16 & 85  & 20 \\
55 & 17 & 90  & 24 \\
60 & 19 & 95  & 17 \\
65 & 19 & 100 & 17 \\
\bottomrule
\end{tabular}
\end{minipage}
\end{table}

\subsection{Cross-Dialect Portability}

To test whether the system produces semantically equivalent QIR regardless of input dialect, we translated 14 matched Catalyst--Quake pairs through the deterministic path. All 14 achieve gate-level match; for the 11 pairs with TVD computable on both sides (probs mode), Catalyst-side TVD averages 99.4\% and Quake-side 98.5\%. The two known-failing random circuits again produce TVD near 0\% despite correct gate counts; GHZ-100 is gate-matched but excluded from TVD comparison.

\subsection{Unseen Dialect Translation}
\label{result_unseen_dialect}
To evaluate extensibility beyond the two registered dialects, we tested the agentic path on three circuits written in a custom FTQC dialect encoding logical qubits via the Steane [[7,1,3]] code, using operations (\texttt{ftqc.init\_zero}, \texttt{ftqc.logical\_h}) absent from the deterministic parser's gate table. The system identifies the \texttt{ftqc} namespace as unknown and routes to the agentic path, equipping the agent with web search and simulator discovery tools.

Table~\ref{tab:e7} reports results across all six models. Llama~3.1 (both variants), Gemma~4, and GPT-OSS-20B all produce valid QIR for all three circuits on the first iteration (6--23~s avg., Table~\ref{tab:e7}). Code Llama~13B succeeds on simpler circuits but fails on the three-qubit one; 34B fails entirely, both emitting tool-call JSON because their tool-use tuning overrides the prompt's request for direct output, and the guardrail cannot recover content never generated.

\begin{table*}[t]
\centering
\caption{Unseen dialect translation across six models. Success rate is out of 3 runs per circuit. No MLIR backend is yet integrated for FTQC, so TVD-based testing is not applicable. Average time is per successful translation, except Code Llama 34B (0/9 successes), whose column reports mean wall-clock time per attempt.}
\label{tab:e7}
\renewcommand{\arraystretch}{1.0}
\begin{tabular}{@{}llccccr@{}}
\toprule
\textbf{Model} & \textbf{Params} & \textbf{1Q Hadamard} & \textbf{2Q Bell} & \textbf{3Q Grover} & \textbf{Total} & \textbf{Avg Time (s)} \\
\midrule
Llama 3.1        & 8B   & 3/3 & 3/3 & 3/3 & \textbf{9/9} & 6.1 \\
Llama 3.1        & 70B  & 3/3 & 3/3 & 3/3 & \textbf{9/9} & 11.9 \\
Code Llama       & 13B  & 3/3 & 3/3 & 0/3 & 6/9 & 14.8 \\
Code Llama       & 34B  & 0/3 & 0/3 & 0/3 & 0/9 & 46.5 \\
GPT-OSS-20B      & 20B  & 3/3 & 3/3 & 3/3 & \textbf{9/9} & 22.9 \\
Gemma 4          & 31B  & 3/3 & 3/3 & 3/3 & \textbf{9/9} & 7.7 \\
\bottomrule
\end{tabular}
\end{table*}
The generated QIR from successful translations is a logical-level mapping: the LLM identifies the gate semantics of the FTQC operations but does not expand them into the physical Steane encoding, which requires a 12-gate encoding circuit per logical qubit (three Hadamards and nine CNOTs) plus seven transversal gates per logical operation. The Catalyst and Quake Steane benchmarks in Table~\ref{tab:circuits} are precisely this expansion for a logical Hadamard: 12 encoding gates followed by a transversal Hadamard on all seven physical qubits, giving the 10 Hadamards and 9 CNOTs listed there. The distinction between logical and physical QIR is inherent to fault-tolerant quantum computing, where the encoding is typically the responsibility of the backend runtime rather than the IR.

\section{Discussion}\label{sec:discussion}

\textbf{Why runtime agentic translation, and not an offline developer tool?} One could argue a deterministic pass is a one-time cost, better written offline than run in a slower runtime loop (1--26~ms at 100\% correctness vs.\ 6.1--46.5~s per translation, Table~\ref{tab:e7}). Our forced-agentic result agrees for some models, though Gemma~4 matches Path~1 accuracy outright, so the case for Path~1 rests on speed rather than a correctness gap only agentic translation could close. Algorithm~\ref{alg:translation} still routes around the agentic path once a dialect's grammar is known. The agentic path earns its place where a deterministic pass does not yet exist: at first contact with a new dialect such as FTQC, where its output can seed the pass written from it later, and in Path~3, for edge cases any hand-written pass eventually misses.

\textbf{Position in a quantum-HPC stack.} Cross-platform execution engines such as QIR-EE~\cite{wong2025qiree} consume QIR to dispatch circuits across heterogeneous backends~\cite{beck2024integrating}. We contribute the stage directly upstream of that dispatch layer: a dialect-agnostic path to QIR and a testing gate that checks the result by execution before dispatch.

\textbf{Context engineering vs.\ RAG.} Injecting the gate mapping table and QIR template ($\sim$2,000 tokens) eliminated the retrieval failures 8B models exhibited under RAG, while still fitting small-model context windows.

\textbf{Architecture vs.\ size.} On these simple FTQC circuits, the results suggest architecture matters more than size: four models spanning 8B--70B (Llama~3.1 8B/70B, GPT-OSS-20B, Gemma~4) all reach 9/9, while Code~Llama 13B/34B score 6/9 and 0/9, the larger model faring \emph{worse} for the tool-use reason given in Section~\ref{sec:results}. Tool-use tuning rather than parameter count separates the two groups. The logical-vs-physical QIR gap reflects domain knowledge absent from current training data, addressable via few-shot QEC examples or fine-tuning.

\textbf{Toward testing unseen dialects without a registered simulator.} Stage~3 (Section~\ref{sec:testing}) requires a registered MLIR-side backend, absent for a never-seen dialect, so unseen-dialect translations are checked only for QIR-side executability. We propose two unimplemented directions: a polynomial-time stabilizer simulator~\cite{nielsen2010quantum} for our Clifford-dominated FTQC circuits (Steane $[[7,1,3]]$), and a restricted state-vector fallback over agent-claimed logical semantics for non-Clifford operations; the stabilizer route seems more promising near-term.

\textbf{Limitations.}\label{sec:limitations} Simulation-based testing requires $2^N$ memory, so circuits above 30 qubits can only be gate-checked. As circuit complexity grows, simulation alone will not scale, and pairing it with formal equivalence checking (e.g., symQV~\cite{bauer2023symqv}) becomes necessary. A small number of mid-circuit-measurement circuits produce correct gate counts but divergent simulation distributions, because the MLIR-side executor does not fully reconstruct classical feed-forward logic (QIR-side execution is correct), a backend reconstruction limitation rather than a translation error. The autodiff gradient circuit is gate-checked only, since it returns an expectation value and its gradient rather than sampled measurements, so distribution-based testing does not apply. For the FTQC dialect, no MLIR backend exists, limiting testing to structural inspection; the preceding paragraph discusses candidate approaches for closing this gap. Evaluation across additional MLIR dialects (OpenPulse, Quil, Stim) would strengthen generalizability claims.

\section{Conclusion}\label{sec:conclusion}

We presented a hybrid system for MLIR-to-QIR translation that integrates deterministic compiler techniques with LLM-based agentic translation and iterative repair, routing each circuit through the fastest reliable of three paths and checking the result by executing source MLIR and target QIR on independent simulators.

On a 41-circuit benchmark spanning Catalyst, Quake, and the unseen FTQC dialect, with GHZ circuits scaling from 5 to 100 qubits, the deterministic path achieved 100\% gate-level correctness on all 70 pairs and average TVD similarity above 95\% across simulable circuits; 14 matched Catalyst--Quake pairs confirmed gate-level equivalence. On FTQC, the agentic path produced valid logical-level QIR for all three test circuits using four of six models tested, with no dialect-specific parser change.

Future work will extend the system to additional MLIR dialects and combine formal verification with simulation-based TVD checking. Concretely: (1) add a Catalyst counterpart for Grover/Shor via Qiskit-to-PennyLane conversion, plus quantum machine learning workloads from PennyLane/CUDA-Q; (2) broaden Path~3 stress-testing beyond the mid-circuit-measurement failure class to isolate cases where repair is achievable; (3) implement the stabilizer-simulation strategy of Section~\ref{sec:discussion}; and (4) identify which model properties (architecture, tool-use tuning, scale) let forced-agentic translation match Path~1 accuracy (Gemma~4 does, Llama~3.1 does not).

\section*{Acknowledgment}

This work was supported by the U.S. Department of Energy, Office of Science under Contract No. DE-AC05-00OR22725, with funding from the Office of Advanced Scientific Computing Research's Accelerated Research in Quantum Computing Program's Modular and Error-Aware Software Stack for Heterogeneous Quantum Computing Ecosystems (MACH-Q) project.

\bibliographystyle{IEEEtran}
\bibliography{references}

\clearpage
\appendices
\section{Artifact Description (AD)}\label{app:ad}

\subsection{Overview of Contributions and Artifacts}
We provide one repository containing the translation system, the benchmark circuit suite, and the experiment scripts used to produce every table and figure reported in this paper.

\subsubsection{Paper's Main Contributions}
\textbf{C1}: A three-path translation strategy that dynamically selects between deterministic, fully agentic, and hybrid repair modes based on dialect recognition and testing outcomes.
\textbf{C2}: A dual-backend testing pipeline that independently executes source MLIR and generated QIR on separate simulators and compares output distributions via TVD.
\textbf{C3}: A guardrail-assembler pattern that recovers valid QIR from structurally incomplete LLM output.
\textbf{C4}: A context engineering approach that injects gate mappings and QIR templates directly into agent prompts.

\subsubsection{Computational Artifacts}
\textbf{$A_1$}: Agentic MLIR-to-QIR Translator, publicly available at \url{https://github.com/ORNL-QCI/Agentic-MLIR_QIR}.

\begin{table}[h]
\centering
\scriptsize
\begin{tabular}{@{}lcp{4.2cm}@{}}
\toprule
Artifact ID & Contrib.\ & Related Paper Elements \\
\midrule
$A_1$ & C1--C4 & Tables~\ref{tab:circuits}, \ref{tab:e1}, \ref{tab:e2}, \ref{tab:e7}; Figs.~\ref{fig:architecture}, \ref{fig:testing} \\
\bottomrule
\end{tabular}
\end{table}

\subsection{Artifact Identification}
\subsubsection{Computational Artifact $A_1$}

\paragraph{Relation To Contributions} $A_1$ is the complete system described in the paper: \texttt{translate.py} (the CLI entry point), the \texttt{agentic\_mlir\_qir/} package (deterministic parsers and generators, LLM agents, dual-backend testing), the benchmark circuit suite under \texttt{example/}, and the experiment drivers under \texttt{experiments/}. Reproducing any experiment exercises all four contributions: dialect routing (C1), TVD-based testing (C2), guardrail-assembler recovery on agentic output (C3), and the context-engineered agent prompts (C4).

\paragraph{Expected Results} Running the deterministic path over the benchmark suite should reproduce Table~\ref{tab:circuits}'s circuit counts and Table~\ref{tab:e1}'s near-100\% gate match, with TVD close to the reported averages (Catalyst 96.9\%/99.2\%, Quake 97.5\%/98.5\%, shots/probs); shots-mode TVD varies run to run from sampling stochasticity. The GHZ scaling sweep should reproduce Table~\ref{tab:e2}'s 100\% gate match from 5 to 100 qubits. The unseen-dialect (FTQC) experiment should reproduce Table~\ref{tab:e7}'s per-model success counts for Llama~3.1 (8B/70B), Gemma~4, and GPT-OSS-20B (9/9); agentic results depend on the LLM provider's serving conditions and may vary between runs.

\paragraph{Expected Reproduction Time (in Minutes)} Setup (\texttt{scripts/setup.sh}) completes in a few minutes without downloading any LLM weights. Deterministic experiments: E1 (Table~\ref{tab:e1}) $\sim$15--30~min, E2 (Table~\ref{tab:e2}) $\sim$10--20~min, E3 (cross-dialect portability) $\sim$10--20~min. The agentic experiment (Table~\ref{tab:e7}) additionally requires an LLM backend (local Ollama or a HuggingFace token) and takes 2--4~hours across all six models, dominated by LLM inference and 70B model loading; a single model on the three paper-reported circuits (9 translations) completes in a few minutes.

\paragraph{Artifact Setup (incl. Inputs)}
\textit{Hardware.} Paper results were collected on an NVIDIA RTX 6000 Ada GPU (48~GB VRAM), Intel Xeon w5-2465X, 256~GB RAM, Ubuntu 22.04. The deterministic path (Tables~\ref{tab:circuits}, \ref{tab:e1}, \ref{tab:e2}) needs no GPU; the agentic path (Table~\ref{tab:e7}) needs a local GPU for Ollama, or none at all against a HuggingFace-hosted model.

\textit{Software.} Python~3.12, PennyLane~0.44.0, PennyLane-Catalyst~0.14.0, \texttt{qir-runner}~0.9.1, CrewAI~1.10.0, and (optional, Linux-only, enables OpenQASM input) CUDA-Q~0.13. All versions are pinned in \texttt{requirements.txt} and \texttt{pyproject.toml}.

\textit{Datasets/Inputs.} The benchmark circuit suite (Table~\ref{tab:circuits}) ships with the repository under \texttt{example/catalyst\_mlir/}, \texttt{example/quake\_mlir/}, and \texttt{example/ftqc\_mlir/}, plus the Grover and Shor circuits under \texttt{example/qiskit\_algorithms/}.

\textit{Installation and Deployment.}
\begin{verbatim}
bash scripts/setup.sh
source venv/bin/activate
\end{verbatim}
This creates a virtual environment, installs \texttt{requirements.txt}, and bootstraps a \texttt{.env} file for the HuggingFace token used by the agentic path (only needed for Table~\ref{tab:e7}). The deterministic path requires no token.

\paragraph{Artifact Execution}
\begin{verbatim}
# Smoke test (~2 min, no LLM required)
bash experiments/run_smoke_test.sh

# E1 -- correctness (Table V)
bash experiments/run_e1_correctness.sh

# E2 -- GHZ scalability (Table VI)
bash experiments/run_e2_scalability.sh

# E3 -- cross-dialect portability
bash experiments/run_e3_cross_dialect.sh

# E4 -- unseen-dialect FTQC translation
# (Table VII; needs HF_TOKEN or Ollama)
bash experiments/run_e4_unseen_dialect.sh
\end{verbatim}
A single circuit can also be translated directly, e.g.\ \texttt{python translate.py example/catalyst\_mlir/code\_bell.mlir --json}. An interactive demo, \texttt{streamlit run agentic\_mlir\_qir/ui/app.py}, is also included but is not required to reproduce any reported result.

\paragraph{Artifact Analysis (incl. Outputs)} Each script writes newline-delimited JSON to \texttt{experiments/results/<experiment>/*.jsonl}, one record per translated circuit with gate-match, TVD, and timing fields. \texttt{experiments/analyze\_results.py} aggregates these into the summary tables reported in the paper. Pre-computed summaries from the paper's own runs are kept under \texttt{experiments/paper\_results/} for direct comparison against a fresh reproduction.

\section{Artifact Evaluation (AE)}
\subsection{Computational Artifact $A_1$}

\paragraph{Artifact Setup (incl. Inputs)} Artifact $A_1$ will be available on GitHub (\url{https://github.com/ORNL-QCI/Agentic-MLIR_QIR}). Hardware, software, and installation match Appendix~\ref{app:ad}.
\begin{verbatim}
git clone <repo_url>
cd Agentic-MLIR_QIR
bash scripts/setup.sh
source venv/bin/activate
\end{verbatim}

\paragraph{Artifact Execution} Run the smoke test, then any experiment driver; each is independent of the others.
\begin{verbatim}
bash experiments/run_smoke_test.sh
bash experiments/run_e1_correctness.sh
bash experiments/run_e2_scalability.sh
bash experiments/run_e3_cross_dialect.sh
bash experiments/run_e4_unseen_dialect.sh
\end{verbatim}
The last script needs \texttt{HF\_TOKEN} set in \texttt{.env}, or a local Ollama with the models listed in \texttt{experiments/README.md}.

\paragraph{Artifact Analysis (incl. Outputs)} Compare \texttt{experiments/results/} against \texttt{experiments/paper\_results/} via \texttt{experiments/analyze\_results.py}. The deterministic path (Tables~\ref{tab:circuits}, \ref{tab:e1}, \ref{tab:e2}) has no stochastic component beyond shots-mode sampling and should match exactly; Table~\ref{tab:e7}'s agentic results vary between runs since LLM output is not deterministic.

\end{document}